\documentstyle[12pt]{article}
\newlength{\bredde}
\def\slash#1{\settowidth{\bredde}{$#1$}\ifmmode\,\raisebox{.15ex}{/}
\hspace*{-\bredde} #1\else$\,\raisebox{.15ex}{/}\hspace*{-\bredde} #1$\fi}
\newcommand{\beq}{\begin{equation}}
\newcommand{\eeq}{\end{equation}}
\newcommand{\bea}{\begin{eqnarray}}
\newcommand{\eea}{\end{eqnarray}}

\def\gtwid{\raise.3ex\hbox{$>$\kern-.75em\lower1ex\hbox{$\sim$}}}
\def\ltwid{\raise.3ex\hbox{$<$\kern-.75em\lower1ex\hbox{$\sim$}}}
\begin{document}
%\begin{Ntitlepage}
%\docnum{CERN--TH-7222/94}
%\vspace*{1cm}
\topmargin -0.8cm
\oddsidemargin -0.8cm
\evensidemargin -0.8cm
\title{\Large{
Screening and Deconfinement of Sources in Finite Temperature
SU(2) Lattice Gauge Theory}}

\vspace{0.5cm}

\author{{\sc P.H. Damgaard and M. Hasenbusch} \\
CERN -- Geneva \\
}
\maketitle
\vfill
\begin{abstract} Deconfinement and screening of higher-representation
sources in finite-temperature $SU(2)$ lattice gauge theory is
investigated by both analytical and numerical means. The effective
Polyakov-line action at strong coupling is simulated by an efficient
cluster-updating Monte Carlo algorithm for the case of $d\!=\!4$ dimensions.
The results compare very favourably with an improved mean-field solution.
The limit $d\!\to\!\infty$ of the $SU(2)$ theory is shown to be highly
singular as far as critical behaviour is concerned. In that limit
the leading amplitudes of higher representation Polyakov lines vanish
at strong coupling, and subleading exponents become dominant. Each of
the higher-representation sources then effectively carry with them their
own critical exponents. 
\end{abstract}
\vfill
\vspace{4.5cm}
\begin{flushleft}
hep-lat/9404008 \\
CERN--TH-7222/94 \\
April 1994
\end{flushleft}
%\end{Ntitlepage}
%\thispagestyle{empty}
\newpage
%\phantom{}
%\vfill
%\eject

%\setcounter{page}{1}

For non-Abelian gauge theories one can define
a static potential between matter sources transforming as arbitrary
irreducible representations of the gauge group, from now on taken to be
$SU(N)$. This static potential
between two infinitely heavy sources is believed to depend crucially
on the manner in which the chosen representation behaves under
transformations restricted to the center $Z(N)$ of the gauge group.
Representations that are insensitive to $Z(N)$ transformations
should yield a screened potential, while those sensitive to these
transformations should yield a confining potential. This is the standard
picture of confinement and screening in non-Abelian gauge theories, 
dating back almost twenty years (see, $e.g.$, ref. \cite{Mack}). 

With numerical simulations these ideas can be tested in the context of
lattice gauge theories. Indeed, one can go further and investigate in more
details the dynamics behind these screening and confining mechanisms.
In particular, one can measure the relevant distance scales, those that
separate ``short distances'' (essentially perturbative physics, on
account of asymptotic freedom) from ``long distances'' (a confining string
between two irreducible sources transforming non-trivially under
$Z(N)$, a screened potential between those transforming
trivially). The dynamics of the intermediate region has
been found to be, in many respects, quite rich. It has been
observed that even representations transforming trivially under the
center group could feel a linearly rising potential (with a slope
different from the string tension of the fundamental representation)
at intermediate distances \cite{Ambjorn}. At a certain range of distance
scales all representations appear to carry with them their own dynamics.
In the limit of an infinite number of colours, factorization is sufficient
to show that {\em all} irreducible representations are confined by
a linearly rising potential if the fundamental representation is, with 
string tensions that depend on the
representations \cite{Greensite}. Essentially, the intermediate distance
region in which a non-zero string tension exists for all representations
grows with $N$, the number of colours, reaching infinity as $N \to \infty$.

These results ought to have some bearing on the physics of 
finite-temperature gauge theories as well. Indeed, some very simple numerical
simulations have shown that just as, for example, adjoint sources may feel
a linearly rising potential at intermediate distances, such sources
appear to ``deconfine'' at precisely the same critical temperature $T_c$
at which fundamental sources deconfine \cite{me1}.\footnote{Even 
representations
transforming non-trivially under $Z(N)$ could in principle deconfine at
temperatures different from the temperature at which the fundamental
representation deconfines. This has not been observed, however 
\cite{Gross1}.} Since, for finite $N$,
adjoint sources are not genuinely confined out to arbitrarily large
distances this is of course only to be understood at the qualitative level.

For gauge theories with continuous deconfinement phase transitions one
obvious question concerns the {\em critical behavior} of the Polyakov lines
corresponding to sources of arbitrary representations. Again, representations
transforming trivially under $Z(N)$ should not be able to serve as order
parameters for the phase transitions, while those sensitive to $Z(N)$
should, since it is precisely a global
$Z(N)$ symmetry which is broken at the deconfinement phase transition
\cite{Polyakov}. Universality arguments would, for such continuous phase
transitions, place the finite-temperature $SU(N)$ gauge theory in the
universality class of globally $Z(N)$ invariant spin systems with 
short-range interactions \cite{Svetitsky}. The analogue of the spin
operator would be the Polyakov line in the fundamental representation.
What about Polyakov lines corresponding to traces taken in higher
representations of $SU(N)$? There appears to be no room for independent
exponents for these higher representations from the spin-system point
of view, since there is simply no obvious analogue of the ``internal''
$SU(N)$ degrees of freedom in the $Z(N)$ fixed-point language. This would
indicate that all higher representations sensitive to the center group should
be equivalent order parameters with exactly the same critical behavior
as the fundamental representation. Physically, this would also be in
agreement with the screening argument according to which all higher
representations sensitive to $Z(N)$ eventually, for large enough
distances, should be colour screened as far down as possible ($i.e.$
down to the charge in the fundamental representation). On the other hand,
since strings with representation dependent string tensions {\em do} form
at some intermediate distance scales for {\em all} representations, it is
not quite obvious how much of the screening mechanism will be
observable at standard present-day lattice sizes used for Monte Carlo
simulations. The clear change in behavior of the adjoint Polyakov line
at the deconfinement ``phase transition''(there is of course no genuine
transition in a finite volume) as measured on small lattices \cite{me1}
is already one indication that there may be difficulties with numerical
investigations of this problem. With the same level of statistics
(and the same lattice sizes) that are used routinely to confirm the
universality arguments based on the fundamental Polyakov line, 
a surprisingly different behaviour was found for the higher representations
of $SU(2)$ lattice gauge theory in (3+1) dimensions\cite{me2,Redlich}.
(For the continuous deconfinement transitions of $SU(2)$ and $SU(3)$
lattice gauge theories in (2+1) dimensions, see ref. 
\cite{Jesper1,Jesper2}). These numerical simulations indicated that sources
of higher representations that were sensitive to $Z(N)$ would correspond
to {\em different} magnetization exponents, one exponent for each
representation. But these results could all be criticized 
\cite{me2,Jesper1} on the grounds that they {\em also} seemed
to indicate critical behavior for Polyakov lines that simply could
not be order parameters for the transition, those of transforming
trivially under $Z(N)$.
Indeed, in a Monte Carlo study of (3+1)-dimensional $SU(2)$ lattice gauge by 
Kiskis \cite{Kiskis1} the expected behavior (adjoint Polyakov line 
non-vanishing across the transition, the isospin 3/2 Polyakov line
behaving like the fundamental) was eventually extracted {\em very close
to the finite-volume ``critical point''} $T_c$. Some of the numerical
difficulties involved are discussed in ref. \cite{Fingberg}.\footnote{For 
a recent discussion of the value of the adjoint Polyakov line at the phase
transition point, see also ref. \cite{Kiskis2}.}

It is important to realize that all of these issues can be addressed
even in the strong-coupling region of the lattice theory. In fact, in this
regime the universality arguments are even strengthened
(since the effective Polyakov-line interactions can be shown explicitly
to be short-ranged \cite{Polonyi,Green}), and the question of the critical
behavior of higher-representation
sources near the phase transition point is as meaningful in the 
strong-coupling regime as near the continuum limit. The advantage of
going to the strong-coupling regime is of
course that the question here can be studied in a much simplified setting
which still captures all the essentials. The leading-order effective 
Polyakov-line action reads, with $Tr_1 W$ indicating the trace in the
fundamental representation \cite{Green}:
\beq
S_{eff}[W] = \frac{1}{2}J \sum_{x,j} \left\{Tr_1 W(x) Tr_1 W^\dagger(x+j) 
+ Tr_1 W^\dagger(x) Tr_1 W(x+j)\right\}~.
\eeq
Here the sum on $j$ runs over nearest neighbours.
The effective coupling $J$ is related to the gauge coupling $g$ and
$N_\tau$, the number of time-like links in the compactified temporal
direction. To lowest order, for $SU(2)$, it is 
\beq
J(g,N_\tau) = \left(\frac{I_2(4/g^2)}{I_1(4/g^2)}\right)^{N_{\tau}} ~,
\eeq
with $I_n$ indicating the $n$th order modified Bessel function.
For $SU(2)$ we will use a notation
in which $Tr_n W$ means the trace taken in the representation of isospin
$n/2$. $Tr_2 W$ is thus the trace in the adjoint representation, etc. 
Higher orders in the 
expansion (1) (and corrections to the effective coupling (2)) can be
computed in a systematic expansion \cite{Gross2}, but we will not need
these corrections for the present purpose. The effective action (1)
becomes asymptotically exact in the strong coupling limit. 

For what follows, it is useful 
to write
the effective Polyakov-line action (1) for $SU(2)$ in terms of a new
variable $\Phi(x) \equiv \frac{1}{2} Tr_1 W(x)$. The partition function
then takes the following form:
\beq
\cal{Z} = \int_{-1}^{1}[d\Phi] \exp
\left[ 4J\sum_{x,j} \Phi(x)\Phi(x+j) + \sum_x \tilde{V}[\Phi^2] \right] ~,
\eeq
with a local potential $
\tilde{V}[\Phi^2] = \frac{1}{2}\ln\left[1 - \Phi(x)^2\right]$.

There are two simple limiting cases in which the effective Polyakov-line
action (1) can be solved exactly. One is the large-$N$ limit \cite{me3},
where the deconfinement phase transition turns out to be of first order
(in agreement with large-$N$ reduction arguments based directly on
the full gauge theory \cite{Gocksch}), and where universality arguments 
hence cannot be addressed.\footnote{Still, the large-$N$ solution does
display a number of interesting features such as the simultaneous
deconfinement of all higher representations at the transition temperature
$T_c$, independently of whether these transformations transform trivially
under the center symmetry (in this case $U(1)$) or not. All representations
of the Polyakov line are hence equally good order parameters in this special
case, and all display a discontinuous jump at the phase transition.} The 
other exactly solvable case is the mean-field limit in
which $d \to \infty$, with $d$ being the number of spatial dimensions
\cite{me2}. In this limit one finds a genuine second-order critical point 
for the gauge group $SU(2)$ at a critical coupling $J_c \to 0$ as 
$d \to \infty$. For $J > J_c$ all higher-representation 
expectation values $\langle Tr_n W \rangle$ are non-zero \cite{me2}:
\beq
\langle Tr_n W \rangle ~=~  (n + 1) \frac{I_{n+1}(2a)}{I_1(2a)} ~,
\eeq
with $a = 2dJ\langle Tr_1 W\rangle$ being the self-consistent 
mean-field solution for 
the fundamental representation. Surprisingly, they {\em all} display
non-trivial critical behavior close to $J_c$:
\beq
\langle Tr_n W \rangle ~\sim~ (J - J_c)^{\beta_{n}} ~,
\eeq
where $\beta_n = n/2$. For the fundamental representation this just
corresponds to the mean-field Ising magnetization exponent $\beta_1
= \beta = 1/2$, in complete agreement with the universality arguments.
For the higher representations this new critical behavior is highly
unexpected. In this unphysical but exactly solvable
limit all standard screening arguments appear to break down, and we are
seeing new behaviour which is not predicted by universality.\footnote{One
also finds separate exponents $\delta_n\!=\!3/n$ for the behaviour of
$\langle Tr_n W\rangle\!\sim\!h^{1/\delta_{n}}$ at $J\!=\!J_c$ in a small
magnetic field coupled to $Tr_1 W$.}

This $d\!\to\!\infty$ result (5) is disturbing on several counts. It is
normally assumed that the relevant $Z(2)$ spin system universality class
to which the $SU(2)$ finite-$T$ phase transition should belong (if
continuous) would
display ``classical'' mean field exponents all the way down from $d\!=\! 
\infty$ to the upper critical dimension $d_u$ (in this case with 
$d_u = 4$, the critical behaviour being modified by logarithmic corrections
just at $d= d_u$). At a first glance this might seem to indicate that
the non-trivial behaviour (5) for {\em all} representations should
remain valid for all $d > 4$ ($d\!=\!4$ just being the limiting case), with 
non-trivial critical scaling even
for representations of integer isospin, and with new critical exponents
for all isospin half-integer representations as well. The first conclusion
simply cannot be correct (and we will show explicitly below why the argument
is invalid), because at strong coupling one can compute, for example, the
adjoint Polyakov line and see that to first non-trivial order in
$1/g^2$ it is non-zero. This result should be valid at least up to
the phase transition point, if this critical point lies sufficiently
deep inside the strong-coupling region \cite{Kiskis2}. What about
the {\em odd}-$n$ representations? Could it be that they display new
non-trivial critical behaviour of the kind (5) even at $d\!=\!4$? To 
investigate this question we have first performed a Monte Carlo simulation
of the effective Polyakov-line action for $SU(2)$ in $d\!=\!4$ spatial
dimensions. Again, the advantages of using directly this effective
action instead of the full $SU(2)$ lattice gauge theory are enormous
from a numerical point of view (extracting the critical indices by
conventional means is notoriously difficult for lattice gauge theories
due to the large lattices and high statistics required), and at the same time
the question we are addressing is as urgent in the context of the effective
action (1) as in the full $SU(2)$ gauge theory.

We have simulated the model numerically using a cluster algorithm similar 
to that
proposed in ref. \cite{Brower}. The algorithm  consists of two parts. 
First there is a single cluster update 
\cite{Wolff} of the sign of the field $\Phi$.
 The delete probability is given by
\beq
 p_d = \exp(-4 J (\Phi(x) \Phi(x+j) + |\Phi(x) \Phi(x+j)| ))
\eeq 
Then,
since the cluster update is not ergodic, we have supplemented it with a
standard Metropolis update that allows changes in the absolute value of 
the field $\Phi(x)$.   
The ratio of the number of single cluster updates to the number of Metropolis
sweeps is a free parameter of the algorithm. We have fixed it by the following
rule of thumb: for every Metropolis sweep one performs approximately
(lattice volume)/(average cluster size) single cluster updates.
The step size of the 
Metropolis algorithm was chosen such that the acceptance rate 
was approximately $1/2$.
%%%%forHansGerd
  Ben-Av et al. \cite{friends} implemented
  such a combination of the single cluster update  and a local heat-bath
 update 
  for the $N_\tau =1$ finite temperature $SU(2)$ gauge theory in $3+1$ 
  dimensions. Critical slowing down was drastically reduced compared to 
  the local update procedure.  

We have determined the critical coupling of the model using the fourth order
cumulant $U_1 = 1 - \langle m^4 \rangle\!/(3\langle m^2 \rangle^2)$ 
\cite{Binder}, where $m$ is the 
single lattice average over $\Phi$, $i.e.$ the Polyakov line in the 
fundamental representation.
We have first simulated lattices of sizes $L\!=\!4, 6, 8, 12$ and 16 at 
$4 J_0 = 0.55$, which
was our guess for the critical coupling after some preliminary simulations
with low statistics.
We have then measured 20,000 times, with one measurement being performed 
after two
Metropolis sweeps plus the corresponding number of single cluster updates.
The integrated autocorrelation time of $m$ was $\tau_m \approx 1.7$ in units
of measurement steps for 
all our lattices sizes. 
We have subsequently computed the fourth order cumulant $U_1$ in a 
neighbourhood of the actual simulation
coupling, by reweighting $\langle m^2 \rangle$ and 
$\langle m^4\rangle$ to the correct Boltzmann weight

  \begin{equation}
   \langle m^n \rangle (J) =
     \frac{\sum_i m^n_i \exp((-J+J_0) \tilde{S})}
                     {\sum_i  \exp((-J+J_0) \tilde{S})} ,
  \label{reweight}
  \end{equation}
where $\tilde{S}$ is defined by $S = J\tilde{S} + \sum_x\tilde{V}$.
The resulting curves are plotted in Fig. 1. The crossings of the cumulant
provide estimates of the critical coupling $J_c$. The error should vanish
like $L^{1/\nu}$.
  
Crossings between lines corresponding to $L = 4$ and 6, 6 and 8, 8 and 12,
and, finally, between $L = 12$ and 16 occur at $4 J_{cross} = 0.5507(8),
0.5509(5), 0.5511(3),$ and 0.5507(2), respectively.
  We take the crossings of the Binder cumulant for 
$L=12$ and $L=16$ lattices as our best estimate of the critical coupling,
$i.e, ~4 J_c = 0.5507(2)$. The limiting value of this crossing is fixed
within a given universality class, in this case expected to be the one
of the 4-$d$ Ising model. 
Brezin and Zinn-Justin \cite{Brezin} have argued 
that the effective potential for dimensions $d \ge 4$ at the critical 
point is given by just a $\phi^4$ term. It follows that   
 the fourth-order cumulant should take the value $U = 1-r_2/3 = 0.27052... $
with  $r_2 = \Gamma(1/4)^4/(8 \pi^2) = 2.1884396... $
 at the critical point. 
The values we find for  
the fourth-order cumulant $U_1$ itself at the crossings, 0.367(10), 0.372(15),
0.378(14) and 0.350(20)  
are considerably larger than this theoretical prediction, and we see
no significant trend towards smaller values with increasing lattice sizes.
Kim and Patrascioiu \cite{Kim} find values
for the fourth order cumulant of the 
the 4-$d$ Ising model on similar and larger lattices 
that are consistent with our numerical result. 
Also Binder {\em et al.} \cite{Binderetal} 
estimate a value of the critical cumulant
of the 5-$d$ Ising model that is consistent with the above numbers. 
In their 
theoretical discussion Binder {\em et al.} allow
for a finite mass term in the effective potential at the critical point. 

Other universal quantities can be extracted from the fourth-order Binder
cumulant. Its derivative should scale like 
$dU/dJ \sim L^{1/\nu}$
at the critical coupling. A fit of the data  according
to this equation leads to $\nu = 0.490(6)$ , which is consistent
within two standard deviations with the 4-$d$ Ising value $\nu = 0.5$. 

Next, consider the fourth-order cumulant $U_n$ for higher representations. 
Fig. 2 shows the results for $n\!=\!2$.  
With increasing lattice sizes
the cumulant converges toward $2/3$ for couplings both below and above 
the critical point. The value $U\!=\!2/3 $ signals a finite expectation value 
of the 
observable. We hence clearly see that the Polyakov line in the adjoint 
representation is not an order parameter. It stays finite in both phases. The 
fourth-order cumulant for the $n\!=\!3$ representation takes values close
to $2/3$ in the broken phase and values close to zero in the high temperature
phase; it behaves as an order parameter. But the curves do not 
display crossings 
close to the critical coupling predicted from the cumulant for the fundamental
representation. The curve for $L\!=\!16$ comes close to that of the fundamental
representation, so we might expect that for still larger lattices the cumulant
for the $n\!=\!3$ representation converges towards the fundamental ones, 
and that the 
crossings can then be observed at the critical coupling $J_c$. 
The data for the higher representation were so affected by errors that 
no reliable results for the cumulants could be extracted. 

It is instructive to look at the finite-size behaviour of the Polyakov lines
of different representations {\em at} the critical point. 
Values for $n=1, 2, 3, 4$ and 5 are given in table 
1. The numbers are obtained from reweighting the $4 J_0 = 0.55$ simulations. 
 Asymptotically the value of the odd representations
 should scale down with increasing lattice size like $
 \langle Tr_n W \rangle \sim L^{-\beta/\nu} = L^{-1}$,
while the values for the even representations should converge to a 
finite value. It appears that the values for the even 
representations indeed stabilize at a (very small) finite  value. 
The behaviour of the magnetization of the odd representations are best
visualized by looking at $\langle Tr_n W \rangle L$, which in the asymptotic
scaling regime should give a 
constant. For $n\!=\!1$
this behaviour is nicely seen in the data, for $n\!=\!3$ there appears to be a 
stabilization for the largest lattice sizes, while for $n\!=\!5$ no 
sign of stabilization is visible on the lattices we have considered. 
It thus appears that the higher the representation, the larger lattices
are required to see the correct infinite-volume behaviour. Physically,
this also makes sense in the light of screening considerations. The higher
the representation, the more screening is required to reproduce the
behaviour of the fundamental representation, and the larger
distances one needs to probe in order to see this.

We have also simulated the model for various $J > J_c$ on lattices of 
sizes up to $L\!=\!16 $. 
Here we have typically performed $10,000$ measurements per simulation,
the aim being an approximate determination of the critical 
exponents $\beta_n$ 
directly from the Monte Carlo measurements of the different representations.  
We have not attempted to fit our data to an ansatz including corrections to 
scaling, but have instead defined a $J$-dependent  
``effective''  exponent $\beta^{eff}_n$ by 
\begin{equation} 
\beta^{eff}_n = (J-J_c) \frac{d\langle Tr_n W \rangle/dJ}{\langle Tr_n 
W \rangle}.
\end{equation} 
The derivative of $\langle Tr_n W\rangle$ with respect to $J$ was computed 
from the relation
\begin{equation} 
\frac{d}{dJ}\langle Tr_n W \rangle = \langle (Tr_n W)\cdot \tilde{S}\rangle
- \langle Tr_n W \rangle \langle \tilde{S} \rangle~.
\end{equation} 

The final results are presented in fig. 4 for the odd representations, and 
in fig. 5 for the even ones. We have carefully checked 
the dependence of the result
on the lattice size and included only values which were consistent on the 
two largest lattice sizes considered. It turn out that for the coupling close
to $J_c$ even the $L=16$ lattice was not sufficient to give a stable result
for the $n=5$ representation.\footnote{The curves plotted in these two
figures are improved mean-field predictions. See below.} 

Figs. 4 and 5 demonstrate fairly convincingly that the odd representations
converge toward an effective $\beta_n^{eff} = 1/2$ independent of $n$, while
the even representations converge toward $\beta^{eff}_n = 0$ (as expected
if these representations remain finite at $J_c$). But the plots also 
reveal an interesting phenomenon for larger values of $(J - J_c)/J_c$:
the effective $J$-dependent exponents $\beta^{eff}_n$ quickly
reach a regime of couplings where they are essentially equally spaced, 
growing linearly with $n$. Although they never actually reach the 
mean-field prediction (5), they get quite close, and they certainly
obey the rule $\beta^{eff}_n \sim n\cdot\beta^{eff}_1$ to surprisingly
high accuracy. This is just as for the original observations in the full 
$(3\!+\!1)$-dimensional $SU(2)$ gauge theory \cite{me2,Redlich}. It appears
that this approximate linear relation between the $\beta_n$'s, when 
measured not too close to the critical point, can be viewed as the 
``remnant'' of the $d\!=\!\infty$ solution. It is then only {\em very} close
to the critical point the behaviour changes, and the single critical
exponent $\beta$ emerges for the odd representations, while the even
representations run smoothly across the transition point. We can estimate
this narrow window in the original gauge coupling $4/g^2$ by using the
relation (2). In the case of $N_{\tau} = 2$ the transition occurs at
$4/g_c^2 = 1.6424(4)$. In order to obtain $\beta_eff < 0.625$ (i.e.
25$\%$ above the correct value $\beta=0.5$) 
for the $n\!=\!3$ representation we would have to take $4/g^2 < 1.66$.

While these results may have clarified the situation in the $d\!=\!4$ theory,
we are still left with the surprising $d\!=\!\infty$ results where mean field
theory is believed to be exact. How can they be explained? Consider the
representation of the effective Polyakov-line action given in eq. (3).
This is a $Z(2)$-invariant effective scalar field theory in $d$ dimensions,
as expected on general grounds. But it is a {\em very particular} effective
scalar theory, one that embodies the underlying $SU(2)$ structure (in the
restrictions on the integration interval of $\Phi(x)$, and in the very
special form of the local potential $\tilde{V}[\Phi^2]$, which reflects 
the Haar measure for $SU(2)$).

Since we at this point wish to focus on the $d = \infty$ results,
we can restrict ourselves to ``classical'' mean-field considerations. It
is instructive \cite{Jesper1} to generalize the partition function above
to an arbitrary local potential $V[\Phi^2]$ and relax the limitation
on the integrations over $\Phi(x)$ to be in the interval $[-1,1]$. The
$d = \infty$ solution is then found by considering the single-site
partition function
\beq
\cal{Z}_{SS}~=~\int_{-\infty}^{\infty}[d\Phi]\exp\left[v\Phi 
+ V[\Phi^2]\right] ~,
\eeq
where $v = 4dJ\langle\Phi\rangle$ will be determined by the 
self-consistency solution.
Clearly, for $n$ being any non-negative integer, $\langle\Phi^{2n+1}
\rangle = 0$ unless 
the global $Z(2)$ symmetry is spontaneously broken. Call the critical 
coupling at which this occurs $J_c$. If the phase transition is continuous, 
$\langle\Phi\rangle$ will be small just above $J_c$, and it is meaningful 
to expand in 
$v$ (no matter how large $d$ is taken, once fixed). The result is, for the 
expectation values of the first two non-trivial mean-field moments of 
$\Phi$ \cite{Jesper1}:
\begin{eqnarray}
\langle\Phi^2\rangle &=& \langle\Phi^2\rangle_0 + \frac{1}{2}\left[\langle
\Phi^4\rangle_0 - \left(\langle\Phi^2\rangle_0
\right)^2\right] v^2 + \ldots \cr
\langle\Phi^3\rangle &=& \langle\Phi^4\rangle_0 v + \frac{1}{2}\left[
\frac{1}{3}\langle\Phi^6\rangle_0
- \langle\Phi^2\rangle_0\langle\Phi^4\rangle_0\right] v^3 + \ldots ~,
\end{eqnarray}
where the subscript ``0'' indicates the (constant) expectation value in 
the unbroken phase $J < J_c$.
Higher moments can be worked out analogously, by expanding both
the partition function ${\cal{Z}_{SS}}$ and the unweighted averages 
in powers of $v$. Using the recursion relation $\chi_{n+1} = \chi_n\chi_1
- \chi_{n-1}$ for $SU(2)$ characters, we find the general $d = \infty$
predictions \cite{Jesper1}
\begin{eqnarray}
\langle Tr_2 W\rangle &=& \left[4\langle\Phi^2\rangle_0-1\right] + 
2\left[\langle\Phi^4\rangle_0
- \left(\langle\Phi^2\rangle_0\right)^2\right] v^2 + \ldots \cr
&=& A_2 + B_2 v^2 + \ldots \cr
\langle Tr_3 W\rangle &=& \left[8\langle\Phi^4\rangle_0 - 4\langle
\Phi^2\rangle_0\right] v +
4\left[\frac{1}{3}\langle\Phi^6\rangle_0 - \langle\Phi^2\rangle_0
\langle\Phi^4\rangle_0 + \frac{1}{2}
\langle\Phi^2\rangle_0\right] v^3 + \ldots \cr &=& A_3 v + B_3 v^3 + 
\ldots ~,
\end{eqnarray}
where $A_2, B_2, A_3$ and $B_3$ are (non-universal) constants.
This shows the behaviour expected from universality arguments. The adjoint
Polyakov line will remain non-vanishing across the phase transition at
$J_c$ (and is hence not an order parameter), and the isospin-3/2 
representation scales near $J_c$ as $v$, $i.e.$, as the fundamental
representation. But if we take the particular potential $\tilde{V}[
\Phi^2]$ of eq. (3), and restrict the integration over $\Phi$ to the
interval $[-1,1]$, then devious cancellations occur. One finds 
$\langle\Phi^2\rangle_0 = 1/4$ and $\langle\Phi^4\rangle_0 = 1/8$, leading to
\begin{eqnarray}
\langle Tr_2 W\rangle &=& \frac{1}{8} v^2 + \ldots = 
2d^2J^2\langle\Phi\rangle^2 + \ldots \cr
\langle Tr_3 W\rangle &=& \left[\frac{4}{3}\langle\Phi^6\rangle_0 + 
\frac{3}{8}\right] v^3
+ \ldots ~.
\end{eqnarray}
It is thus suddenly the {\em non-leading} terms in the general expansion
of the Polyakov lines that become important, due to the {\em amplitudes}
of the leading terms vanishing in this limit.
The cancellations required for this phenomenon are actually simple 
consequences of the orthogonality
relations for $SU(2)$ characters, as follows if one performs the
mean field calculation directly in $SU(2)$ language \cite{me2}. They
occur similarly for all higher representations, leading, of course,
eventually to the general $d\!=\!\infty$ solution (5).

It is interesting to compare this result with a general 
renormalization-group analysis by Kiskis \cite{Kiskis1}. Not being restricted
to $d = \infty$, the results of Kiskis can be summarized by
\begin{eqnarray}
\langle Tr_2 W\rangle &=& a_2 + b_2 t^{1-\alpha} + \ldots \cr
\langle Tr_3 W\rangle &=& a_3 t^\beta + b_3 t^{1+\beta} + \ldots ~,
\end{eqnarray}
where $\alpha$ is the usual specific heat critical exponent, and $t$ is
the reduced temperature near the critical point at $T_c$. In the $d\!=\! 
\infty$ limit, these are precisely of the form (12) above {\em provided we
make the identifications}
\beq
1 - \alpha = 2\beta~,~~~~~~~1 + \beta = 3\beta
\eeq
(assuming that the coefficients $B_2,B_3,b_2$ and $b_3$ 
are all non-zero). Solving
these equations, we find $\beta = 1/2$ and $\alpha = 0$, the $d = \infty$
Ising model exponents. With the above qualification, the identities 
(15) appear
to be new scaling relations for the $Z(2)$ fixed point at $d = \infty$, 
imposed by the combined restrictions of mean field theory and the
renormalization group. 

We are now in a better position to understand the $d = \infty$ results.
As shown above, the appearance of
new exponents for each of the odd-$n$ representations in the limit
$d\!=\!\infty$ is due to very delicate cancellations that make the
{\em amplitudes} of the leading terms in the expansion close to the critical
point vanish. Although the same mechanism is responsible for the fact
that also even-$n$ representations display non-trivial
critical behaviour in the $d\!=\!\infty$ theory, that phenomenon is of course
far more difficult to understand from the point of view of physics.   
The even-$n$ Polyakov-line representations simply ought not to be order
parameters for the deconfinement transition, even in the $d\!=\!\infty$ limit,
since such sources should be screened both above and below the
critical point. The resolution of this apparent paradox lies in the fact
that the critical coupling $J_c$ actually {\em vanishes} (like $1/d$) when
$d\!\to\!\infty$, as follows directly from the mean-field solution 
(4).\footnote{This behaviour is not an artifact of the mean-field solution.
It can be checked to hold as well in the exact solution of the  
$N = \infty$ theory \cite{me3}.} In terms of the gauge coupling $g$ this
entails, for fixed $N_\tau$, $g\!\to\!\infty$. Although 
this makes the 
strong-coupling effective Lagrangian analysis more and more accurate,
it also pushes the confinement/deconfinement phase transition right to
the extreme limit $g\!=\!\infty$ where {\em all} sources are ``confined''
($\langle Tr_n W\rangle = 0$ for all $n$ at $g\!=\!\infty$ in the full 
gauge theory
simply as a consequence of
the orthogonality property of the group characters). It is for this simple
reason that the mean-field solution, correctly, predicts critical behaviour
for all representations of $SU(2)$.

The limit $d\!=\!\infty$ of finite-temperature gauge theories is thus in
many respects highly singular. This, together with the Monte Carlo data
presented above for the $d\!=\!4$ $SU(2)$ theory, indicates that the usual
assumption of $d\!=\!\infty$ exponents being valid down to the upper critical
dimension $d_u$ simply fails in this case. Can we understand the 
singular nature
of the $d\!=\!\infty$ limit in an analytical way? As explained above,
there are actually no reasons to doubt that mean field theory predicts
the $d\!=\!\infty$ behaviour correctly. The only resolution would then be
that {\em any} finite dimensionality $d$ should correspond to radically 
different behaviour close to the critical point, $i.e.$, that 
$1/d$-corrections discontinuously should alter the critical indices. To
see whether this is the case, we have considered a slightly improved
mean-field solution of the same effective Polyakov-line action (1). (This
improvement appears to be equivalent to what is known as the 
Bethe-approximation, see, $e.g.$, ref. \cite{bethe}).
Consider a system that consists of $2d\!+\!1$ sites, a central site (C) and 
its $2d$ nearest neighbours (O). The remaining $2d\!-\!1$ sites of the O-sites 
are replaced by an external field $W$, which here is the analogue of the
conventional mean field. Hence the partition function of this
system is given by  
\begin{equation}
Z = \int_{-1}^{+1} d\Phi_C \sqrt{1-\Phi_C^2}
     \prod_O \int_{-1}^{+1} d\Phi_O \sqrt{1-\Phi_O^2}
        exp(4 J (\Phi_C + W)  \Phi_O)
\end{equation}
The integration over the $\phi_O$ fields leads to  
\begin{equation} 
Z \propto \int_{-1}^{+1} d\Phi_C \sqrt{1-\Phi_C^2}
 \left[\frac{I_1(4 J (\Phi_C + W))}{4 J (\Phi_C + W)}\right]^{2d}
\end{equation} 
The remaining one-dimensional integration we have performed numerically.
In order to fix the external field $W$ we require that the magnetization
of the fundamental representation is equal for the central site (C) and its 
neighbours (O).  We solved this condition numerically using the Newton method. 
Expectation values are evaluated on the central site. 

One can readily check that this improved mean field theory coincides with
the conventional mean field theory in the limit $d \to \infty$. It is,
however, expected to be more accurate for finite values of $d$, especially
for non-universal quantities. 
We have solved the self-consistency equations
for $d=4, 8, 16$ and 32.  The value found for the critical coupling $J_c = 
0.5352...$ in 4-$d$ deviates from the Monte Carlo 
value by only $2.8\%$, while standard mean field theory is off by $9.2\%$. 
But a more striking consequence of the improvement is seen in the behaviour
of the {\em even}-n representations, which are now non-vanishing for
all values of the coupling $J \neq 0$. The numerical values at or below
$J_c$, however,
decrease rapidly with $d$. For the adjoint representation it is reduced
by a factor of approximately 2 at $J_c$ when one doubles the dimension,
while for the $n\!=\!4$ representation it drops by almost a factor of 4. In
this fashion the present solution matches the usual mean-field results in
the limit $d\!=\!\infty$.

Since such an apparently minor improvement of the mean-field method produces
this drastic change in behaviour for the even-$n$ representations, it is
worthwhile to understand it better. In fact,
almost {\em any} improvement of the $d\!=\!\infty$ mean-field results are
bound to give qualitatively the right behaviour for the even-$n$
representations. We will here show that, except for
the mean-field limit $d\!=\!\infty$, the expectation values of
Polyakov lines in even-$n$ representations will always
be {\em positive} for all finite $d$ and $J$.
Note that we can rewrite the expectation value $\langle Tr_n W \rangle$ as 
\begin{equation}
 \langle Tr_n W \rangle = \int dS P(S) M_n(S)
\end{equation}
where $M_n(S)$ is the conditional expectation value of $ Tr_n W $
 at 
a site $C$ for fixed sum $S$
 of the fields $\Phi_O$ at the neighboring sites $O$.
  $P(S)$ is the
probability distribution of the sum of the fields $\Phi_O$.
 $M_n(S)$ is given by
\begin{equation}
M_n(S) =  \frac{\int_{-1}^{+1} d\Phi_C 
\exp
\left[ 4J S \Phi_C +  \tilde{V}[\Phi_C^2] \right] Tr_n W}
{\int_{-1}^{+1} d\Phi_C 
\exp 
\left[ 4J S \Phi_C +  \tilde{V}[\Phi_C^2] \right] }
\end{equation}
We have already computed $M_n(S)$, since it is of precisely the same form 
as the mean-field solution of the model: we just
have to take $a = 4 J S$ as argument in eq. (5). For $n$ even $M_n(S)$ 
is a even function, since odd (modified) Bessel-functions are odd functions.
The only real zero of the function is given at $S=0$. For all other real
arguments $M_n(S)$ is strictly positive.

Next consider the conditional probability distributions $p(\Phi_{NO},S)$, where
$\Phi_{NO}$ are the fields on the neighbouring sites of the $O$ sites. 
Note that the computation of 
 $p(\Phi_{NO},S)$ requires only the integration of the $\Phi_O$ fields.
One reads from the definition  that $p(\Phi_{NO},S) > 0$ for $|S|<2d$
 for any configuration
of the $\Phi_{NO}$ with  $|\Phi_{NO}| \le 1$. Hence we also
 have $P(S) > 0$ for $|S|<2d$.

Taking the properties of $M_n(S)$ with $n$ even and $P(S)$ as discussed
above $\langle Tr_n W \rangle$ for n even has to be strictly larger than zero.
Mean-field theory gives zero
 for $J \le J_c$ since $P(S)$ is replaced by $\delta(S)$. Any improvement in 
the mean-field solution that provides a smooth function for
 $P(S)$ will remove the mean-field pathologies that lead to vanishing
expectation values for the even-$n$ representations at and below $J_c$. 
Of course, in the limiting
case $d\!=\!\infty$, the exact $P(S)$ {\em is} a $\delta$-function at
$J\!=\!J_c$, and the naive mean-field results are exact. 

With the improved mean field theory we can finally make a much more accurate
comparison with our $d\!=\!4$ Monte Carlo results. In figs. 4 and 5 we have
thus plotted (as smooth curves) the corresponding predictions for the
$J$-dependent effective exponents $\beta^{eff}_n$ defined as in eq. (8).
Qualitatively the behaviour of the Monte Carlo data is quite well reproduced.
Since it is clear that the conventional mean field results must be
reproduced as $d\to\infty$, it is interesting to see what happens with
these $\beta^{eff}_n$-exponents as $d$ is increased. In fig. 6 we show 
for $n=2$ how
the lowest-order mean-field behaviour is recovered as $d$ grows. One sees
that the window close to $J_c$ where $\beta^{eff}_2$ turns over and starts
deviating from the $d\!=\!\infty$ result $\beta_2\!=\!1$ gets more and
more narrow as $d$ is increased. This behaviour is not restricted
to the adjoint representation. In fig. 7 we show the results for 
all $\beta^{eff}_n$ up to $n\!=\!5$ in $d\!=\!32$ dimensions. The linear
spacing of effective exponents is seen throughout, except for an extremely
narrow interval close to $J_c$. 

Clearly, as $d\!\to\!\infty$ the window in which the conventional
results are reproduced shrinks to zero. In fact, one can easily estimate
from the improved mean-field solution (16) that this window decreases in size
as $1/d$, eventually disappearing at $d\!=\!\infty$. In the more
conventional language, this is the point at which the amplitudes of
the leading terms in the expansion for the Polyakov lines vanish.

\vspace{0.5cm}

%\noindent
{\sc Acknowledgment:} One of us (PHD) would like to thank Jeff Greensite
for helpful discussions.
%%%%forHansGerd 
We would like to thank Hans Gerd Evertz for drawing our attention to ref. 
\cite{friends}. 
%\newpage

\newpage

\begin{table}
%\squeezetable
\caption{
  Magnetizations at the critical coupling 
  $4J_c=0.5507$ for various representations. The number in the 
  first bracket gives the statistical error at the given coupling, while the
  number in the second gives the uncertainty due to the error of the critical
  coupling.
        }
%\label{attc}

\vspace{0.5cm}

\begin{center}
\begin{tabular}{|r|l|l|l|l|l|}
\hline
 L &    $~~\langle Tr_1 W\rangle$  &  $~~~\langle Tr_2 W\rangle$ &  
$~~\langle Tr_3 W\rangle$ &  $~~~\langle Tr_4 W\rangle$    & $~~~\langle 
Tr_5 W\rangle$ \\
\hline
 4 &0.354(2)(1)&0.1605(14)(6)&0.0659(4)(2)&0.0101(6)(1)& 0.0543(3)(0)  \\
 6 &0.249(2)(2)&0.1348(7)(6)&0.0344(3)(2)&0.00802(30)(6)&0.02369(14)(1)\\
 8 &0.194(2)(3)&0.1255(4)(7)&0.0230(2)(3)&0.00693(14)(6)&0.01321(7)(1) \\
12 &0.129(1)(5)&0.1177(2)(6)&0.0136(1)(4)&0.00592(8)(6)& 0.00593(3)(1) \\
16 &0.100(1)(7)&0.1157(1)(7)&0.0102(1)(6)&0.00592(4)(6)& 0.00333(2)(1) \\
\hline
\end{tabular}
\end{center}
\end{table}

\begin{table}
%\squeezetable
\caption{
          Improved mean field theory: $J_c$, and $m_j$ evaluated at $J_c$. 
        }
        % \label{attc}

\vspace{0.5cm}

\begin{center}
\begin{tabular}{|r|c|c|c|}
\hline
 $d$  & $4 J_c$ & $\langle Tr_2 W\rangle$ &    $\langle Tr_4 W\rangle$   \\  
\hline
  4   &  0.5352319055  & 0.07382995665 & 0.00254915596 \\
  8   &  0.2582840308  & 0.03387515014 & 0.00055551163 \\
 16   &  0.1270085358  & 0.01625711255 & 0.00013006341 \\
 32   &  0.0629953883  & 0.00796783664 & 0.00003149409 \\
\hline
\end{tabular}
\end{center}
\end{table}

\begin{center}
\Large{\bf Figure Captions}
\end{center}

\noindent 1.)~The fourth-order Binder cumulant for the fundamental
representation. The crossing determines our best estimate of $J_c$.

\noindent 2.)~Same as fig.1, but for the adjoint source. The convergence
toward $U_2 = 2/3$ indicates that the expectation value of the Polyakov line
in the adjoint representation is non-zero throughout.

\noindent 3.)~The fourth-order Binder cumulant for the $n\!=\!3$ source.
Although the behaviour is radically different from that of the adjoint
source, it still does not show the pattern of fig.1 for the smaller
lattices.

\noindent 4.)~The effective magnetization exponents (8) for odd-$n$. The
$n\!=\!1$ and $n\!=\!3$ representations nicely appear to converge toward
the Ising value of $\beta\!=\!0.5$, while the $n\!=\!5$ representation
only shows the same trend. The drawn curves refer to the improved
mean-field solution discussed at the end of the paper.

\noindent 5.)~Same as fig. 4, but for the even-$n$ representations. The
effective exponent appears to converge toward 0, as expected if these
magentizations remain non-zero at the critical point.

\noindent 6.)~The way the mean-field solution $\beta_2 = 1.0$ is recovered 
in the limit $d\!\to\!\infty$. The region in $(J\!-\!J_c)/J_c$ where
$\beta^{eff}_2$ eventually turns to zero shrinks as $d$ grows, disappearing
in the limit $d\!\to\!\infty$.

\noindent 7.)~The linear spacing of $\beta^{eff}_n$, here for $d\!=\!
32$. 

\vspace{1cm}

\begin{thebibliography}{X}
\bibitem{Mack}G. Mack, DESY preprint 77/58, unpublished; Phys. Lett.
{\bf 78B} (1978) 263.
\bibitem{Ambjorn}C. Bernard, Nucl. Phys. {\bf B219} (1983) 341. 
\newline J. Ambj{\o}rn, P. Olesen and C. Peterson, {\bf B240}[FS12] 
(1984) 189, 533.
\bibitem{Greensite}J. Greensite and M.B. Halpern, Phys. Rev. {\bf D27}
(1983) 2545. 
\bibitem{me1}P.H. Damgaard, Phys. Lett. {\bf B183} (1987) 81.\newline
M. Faber, H. Markum and M. Meinhart, Phys. Rev. {\bf D36} (1987) 632.
\bibitem{Gross1}M. Gross and J.F. Wheater, Phys. Rev. Lett. {\bf 54}
(1985) 389.
\bibitem{Polyakov}A.M. Polyakov, Phys. Lett. {\bf B72} (1978) 477. 
\newline L. Susskind, Phys. Rev. {\bf D20} (1979) 2610.
\bibitem{Svetitsky}B. Svetitsky and L.G. Yaffe, Nucl. Phys.
{\bf B210}[FS6] (1982) 423.
\bibitem{me2}P.H. Damgaard, Phys. Lett. {\bf B194} (1987) 107.
\bibitem{Redlich} K. Redlich and H. Satz, Phys. Lett. {\bf B213} (1988) 191.
\bibitem{Jesper1}J. Christensen and P.H. Damgaard, Nucl. Phys. {\bf B348}
(1991) 226.
\bibitem{Jesper2}J. Christensen and P.H. Damgaard, Nucl. Phys. {\bf B354}
(1991) 339. \newline J. Christensen, G. Thorleifsson, P.H. Damgaard and
J.F. Wheater, Nucl. Phys. {\bf B374} (1992) 225.
\bibitem{Kiskis1}J. Kiskis, Phys. Rev. {\bf D41} (1990) 3204.
\bibitem{Fingberg}J. Fingberg et al., Phys. Lett. {\bf B248} (1990) 347.
\bibitem{Kiskis2}J. Kiskis and P. Vranas, Phys. Rev. {\bf D49} (1994) 528.
\bibitem{Polonyi}J. Polonyi and K. Szlachanyi, Phys. Lett. {\bf 110B}
(1982) 395. \newline J. Bartholomew, D. Hochberg, P.H. Damgaard and
M. Gross, Phys. Lett. {\bf 133B} (1983) 218. \newline M. Ogilvie,
Phys. Rev. Lett. {\bf 52} (1984) 1369. \newline J.-M. Drouffe, 
J. Jurkiewicz and A. Krzywicki, Phys. Rev. {\bf D29} (1984) 2982.
\bibitem{Green}F. Green and F. Karsch, Nucl. Phys. {\bf B238} (1984)
297. 
\bibitem{Gross2}M. Gross and J.F. Wheater, Nucl. Phys. {\bf B240}[FS12]
(1984) 253.
\bibitem{me3}P.H. Damgaard and A. Patkos, Phys. Lett. {\bf 172B} (1986)
369.
\bibitem{Gocksch}A. Gocksch and F. Neri, Phys. Rev. Lett. {\bf 50}
(1983) 1099.
\bibitem{Brower} R.C. Brower and P. Tamayo, Phys. Rev. Lett. {\bf 62}
(1989) 1087.
\bibitem{Wolff} U. Wolff, Phys. Rev. Lett. {\bf 62}
(1989) 361.
%%%%forHansGerd 
\bibitem{friends}  R. Ben-Av, H.G. Evertz, M. Marcu and S. Solomon, 
 Phys.Rev. {\bf D44} (1991) 2953. 
\bibitem{Binder}
      K. Binder, Phys. Rev. Lett {\bf 47} (1981) 693; \newline
      K. Binder, Z. Phys. {\bf B43} (1981) 119.
\bibitem{Kim} J-K. Kim and A. Patrascioiu, Phys. Rev. {\bf D47} (1993) 2588.
\bibitem{Brezin} E. Brezin and J. Zinn-Justin, Nucl. Phys. {\bf B257}
 (1985) , 867.
\bibitem{Binderetal} K. Binder, M. Nauenberg, V. Privman and A. P. Young,
Phys. Rev. {\bf B31} (1985) 31.
\bibitem{bethe} C. Itzykson and J-M Drouffe: {\em Statistical Field Theory}
 Vol. 1, Cambridge University Press (1989). 
\end{thebibliography}
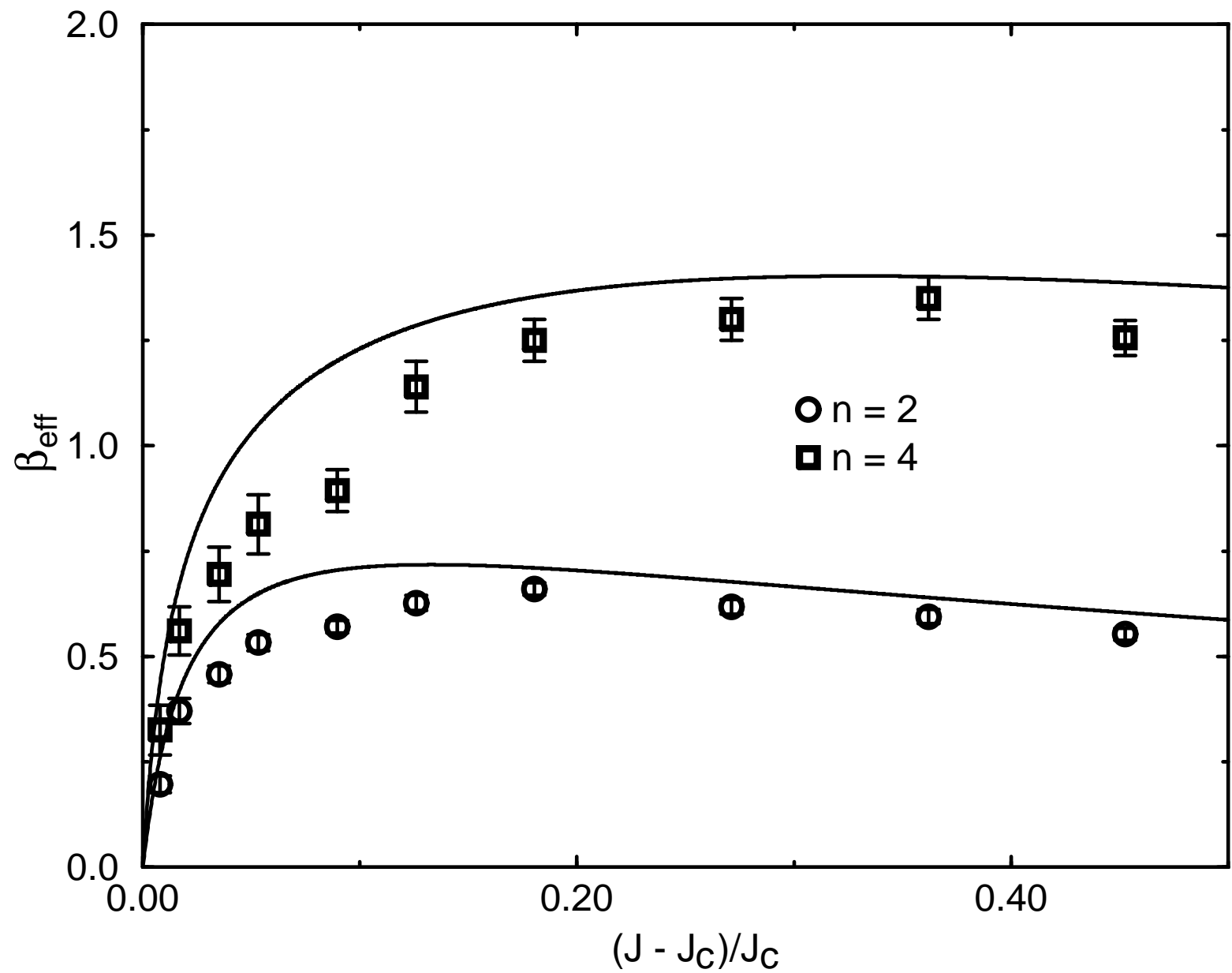
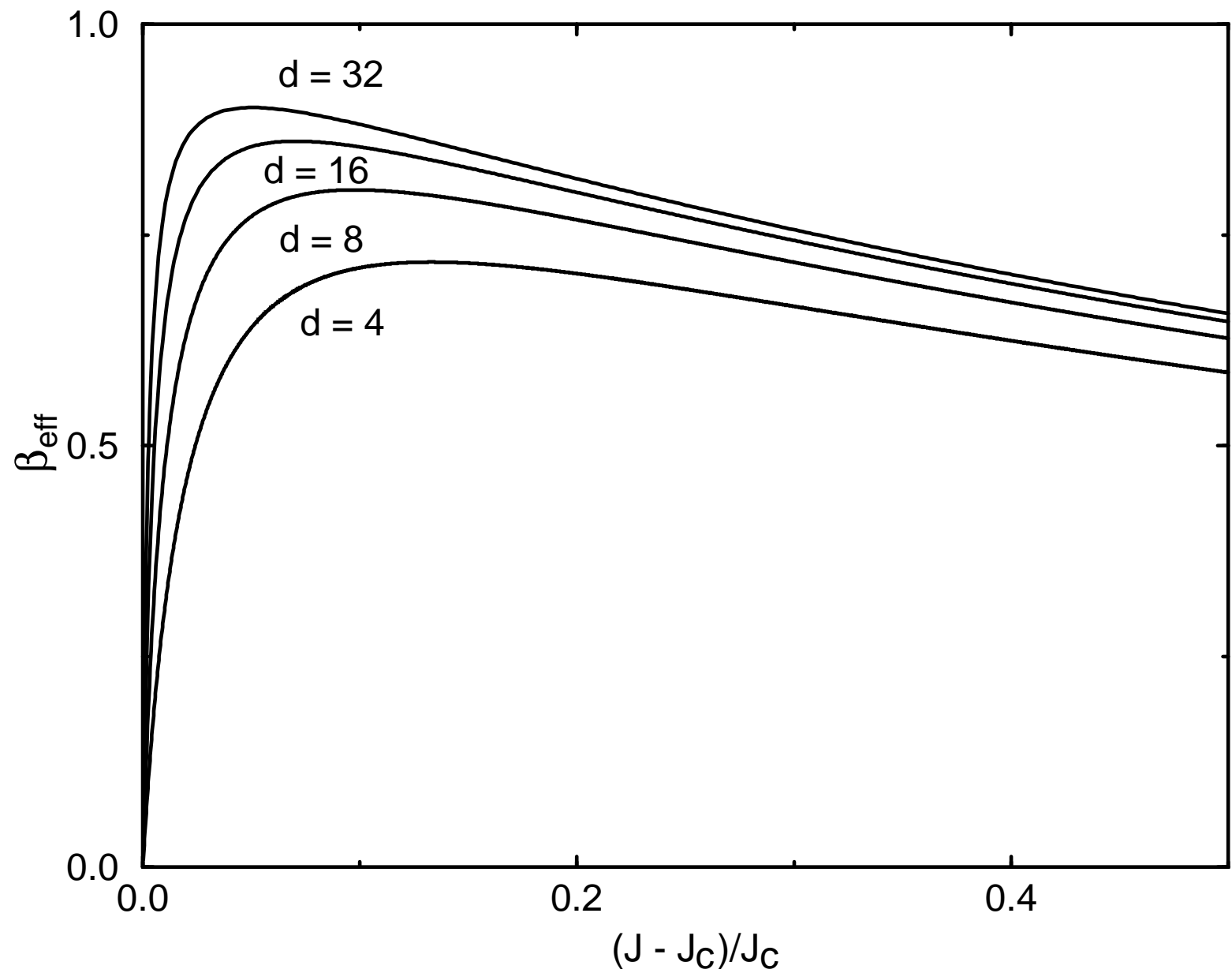
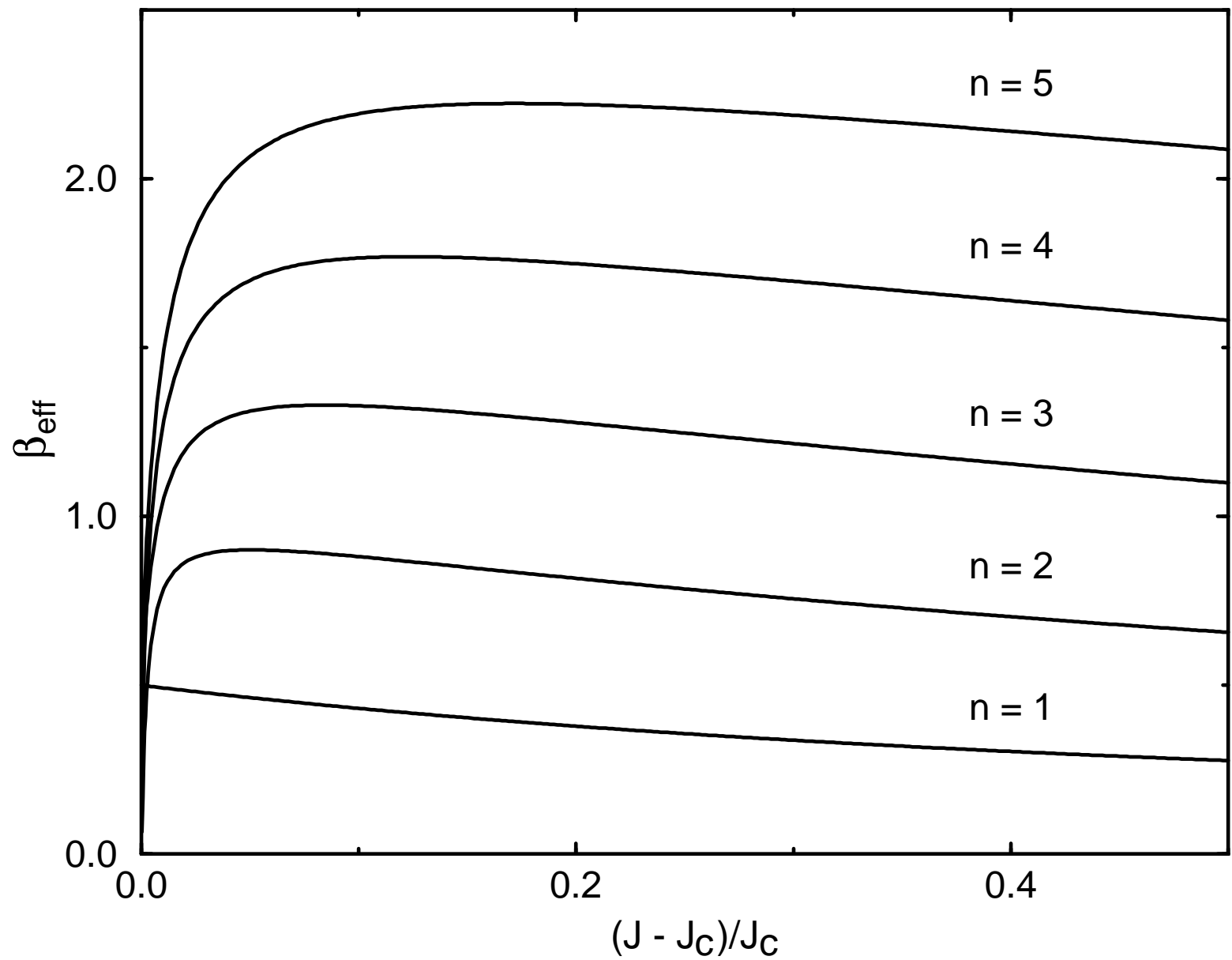
\end{document}